\documentclass{appolb}
\usepackage{graphicx}
\usepackage{hyperref}

\begin{document}

\title{Elliptic flow measurement of $J/\psi$ in PHENIX Run14 Au+Au at $\sqrt{s_{NN}}=200$ GeV
\thanks{Presented at the 29$^{th}$ International Conference on Ultrarelativistic Nucleus-Nucleus Collisions (Quark Matter 2022)}}
\author{Luis Bichon III (for the PHENIX collaboration,\\
\href{https://doi.org/10.5281/zenodo.7430208}{https://doi.org/10.5281/zenodo.7430208})
\address{Department of Physics and Astronomy, Vanderbilt University,
Nashville, TN 37235 USA}}

\maketitle
\begin{abstract}
We obtain the first measurement of $J/\psi$ elliptic flow at RHIC energies in forward rapidity using data from the PHENIX detector and applying an event plane method. The dataset used contains 19 billion events from the PHENIX experiment's Run 14 Au + Au dataset at $\sqrt{s_{NN}}=200$ GeV. PHENIX has measured a $J/\psi$ $v_2$ in a centrality range of $10-60\%$ that is consistent with zero. Taken together with results from LHC the measurement of $v_2$, which is consistent with zero may indicate that $J/\psi$ production by coalescence is not significant at forward rapidity at RHIC energy.
\end{abstract}
  
\section{Introduction}

The QGP has been found to exhibit a nearly perfect fluid behavior~\cite{Heinz_2009}. This behavior manifests itself as strong correlations between particles produced in nuclear collisions. Presently, the detailed interactions of the heavy quarks in the QGP medium are under investigation and, because heavy flavor quarks will have relatively larger masses, they may not be thermalized and flow with the medium. The production of $J/\psi$ in p+p collisions is theoretically well understood because they are produced in hard scattering processes. This feature in addition to their production in hard scattering events in the initial stages of the collision make them ideal probes for testing the properties of the QGP medium. However, in nucleus+nucleus collisions some of the produced $J/\psi$ mesons may be dissolved by the QGP, which may create anisotropies in the observed $J/\psi$ azimuthal distributions due to the different path length in the medium. Additionally, a similar signal may be created if the $J/\psi$ thermalizes inside the medium and follows the pressure gradients as lighter particles do, or the $J/\psi$ may dissociate, and the charm quarks could equilibrate which could lead to $J/\psi$ regeneration. We present a preliminary result for $J/\psi$ $v_2$ using the PHENIX Run14 Au+Au dataset at $\sqrt{s_{NN}}=200$ GeV. 

\section{Data Analysis \& Methodology}
\subsection{Dataset and Detectors}
In this analysis, we use the Run 14 Au+Au Muon Arm dataset at $\sqrt{s_{NN}}=200$ GeV containing 19 billion events. The dimuon decay channel is used to reconstruct candidate $J/\psi$ mesons. The PHENIX experiment has a unique coverage at forward rapidity with muon identification. This in addition to the large dataset of Au+Au collisions collected in 2014 allows for a statistically improved measurement of $J/\psi$ elliptic flow at RHIC energies.

The key detector in this analysis is the Forward Silicon Vertex Detector (FVTX). With the FVTX, an increase in precision vertexing capabilities was added to the muon spectrometers, enabling the rejection of muons from the decay of relatively long-lived particles, the rejection of muons from the decays of relatively long-lived particles, and an additional way of determining the event plane~\cite{Aidala_2014}.

\subsection{Combinatorial Background Subtraction}
To obtain a pure signal for the $J/\psi$ from dimuon mass distributions we employ event-mixing as the standard method of removing the background dimuons. For this event-mixing method, the background is constructed from dimuon pairs of opposite sign, but the single muons come from different events. Mixed event dimuon pairs are only formed if two events have a centrality closer than $5\%$, a $Z$ vertex closer than $0.75$ cm and a event plane angle closer than $\pi/20$ rad. Using events instead of individual dimuons allows us to increase the likelihood that we are using combinatorial background dimuons. A normalization factor must be applied for the background which can be obtained by using the ratio of like-sign pairs from the same event to like-sign pairs from mixed events. The signal is then obtained by the subtraction of the normalized background from the foreground.

\subsection{Fitting the Dimuon Mass Distribution}

In the fitting of the mass distributions, we assume the shape of the $J/\psi$ signal to be a Crystal Ball function, and given the statistical precision of the dataset, we also apply the same shape to the $\Psi(2S)$ to avoid their inclusion in the higher mass $J/\psi$ region. The parameters of the Crystal Ball function are obtained using $J/\psi$ embedded Monte Carlo simulation data. We produce simulated mass distributions for low/high $p_T$ and South/North arm rapidities, fitting the distributions allowing for the function to have free ($\alpha$, $n$, $\bar{x}$, and $\sigma$) parameters. The $J/\psi$ count for each distribution is obtained by the integral of the $J/\psi$ crystal ball function in the fit (see Figure 1).

\begin{figure}[htb]
\centerline{%
\includegraphics[width=\linewidth]{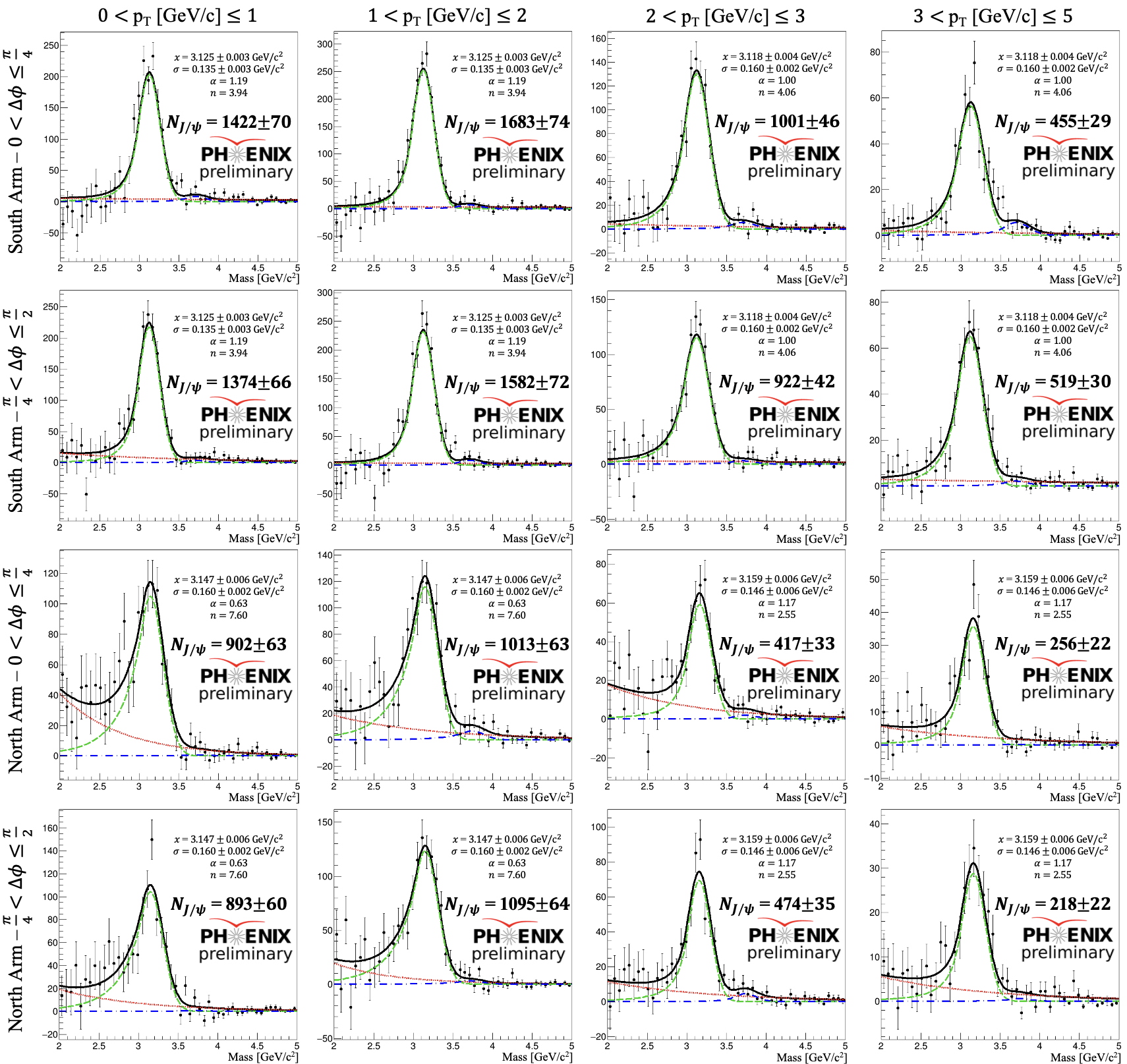}}
\caption{Mass distributions using mixed-event subtraction for the unweighted ``standard" set. These are binned by $p_T$ in each column, and rapidity+$\Delta\phi$ angle for each row. The green/dashed curve is a Crystal Ball fitted to the $J/\psi$ peak, the blue/dashed-dot curve is a Crystal Ball fitted to the $\psi(2S)$ peak, the red/dotted curve is an exponential fitted to the remaining background after subtraction, and the black/solid curve is the total fit.}
\label{Fig:F2H}
\end{figure}

\subsection{Event Plane Method and Measuring $v_2$}

We are primarily using the In/Out ratio method, which is an event plane method~\cite{Poskanzer_1998} that uses the counts of the $J/\psi$ in bins of $\Delta\phi$ to measure $v_2$. The In/Out ratio method splits the distributions into $2$ bins of $\Delta\phi$ one in plane with the event plane and the other out of plane. We measure $v_2$ using this method by looking at the difference between these bins. If there is no preference in either plane, we would observe a flow around zero.

\subsection{Systematic Uncertainties}
The systematic uncertainties are determined by changing various aspects of the analysis. As of this time, we have employed changing the primary detector of the analysis from the FVTX to the Central Arm Spectrometers (CNT), which covers a different pseudorapidity range. We have used a different method for our combinatorial background subtraction, the like-sign method, which constructs the background with dimuon pairs of the same sign ($\mu^+\mu^+$ and $\mu^-\mu^-$) that come from the same event. The uncertainty in the normalization factor in the event-mixing method was also incorporated into the systematic uncertainty. The last systematic uncertainty we consider comes from the mass fitting of the dimuon distribution, where the shape of the continuum distribution was assumed to be an exponential function, and the uncertainty in this assumption can be explored by assuming no continuum contribution in the $J/\psi$ mass region.

\section{Results}

Figure 2 shows the $p_T$-dependent $J/\psi$ $v_2$. The measurement in this analysis for PHENIX Run 14 at forward rapidity in a centrality range of 10 - 60\% is shown in red. The measurement made by STAR at mid-rapidity and in a centrality range of 10-40\% is shown in black. The ALICE result at forward rapidity in a centrality range of 20-40\% is shown in blue. Boxes surrounding the data points represent systematic uncertainties.

PHENIX observes a larger suppression of $J/\psi$ yield in forward rapidity when compared to mid-rapidity. This is contrary to expectations, because effects that dissolve the $J/\psi$ have been determined to be stronger at mid-rapidity~\cite{PhysRevC.84.054912}. To understand this observation we begin by looking into the production of $c\bar{c}$ pairs. The majority of $c\bar{c}$ pairs per event in central collisions at RHIC are produced at mid-rapidity. At LHC energies, less suppression is observed, where many more $c\bar{c}$ pairs per event in central collisions are produced~\cite{Andronic_2018}. To explain this behavior, theoretical models require a contribution of coalescence via a recombination mechanism between charm and anticharm quarks~\cite{Pereira_Da_Costa_2016}. It was found that the strength of this coalescence effect increases with the initial number of produced $c\bar{c}$ pairs relative to the total number of quarks, increasing with the collisions energy.

At LHC energies, a nonzero $v_2$ is observed, this is in line with $J/\psi$ formed by coalescence in the QGP medium, and carrying the azimuthal anisotropy of the system~\cite{2020}. At RHIC energies, STAR has measured $v_2$ that is consistent with zero, but due to limited statistics remains inconclusive~\cite{Adamczyk_2013}. With coalescence being the dominant mechanism for nonzero $J/\psi$ $v_2$ it should follow that systems where fewer $c\bar{c}$ pairs are formed should have a smaller azimuthal anisotropy.

\begin{figure}[htb]
\centerline{%
\includegraphics[width=12.5cm]{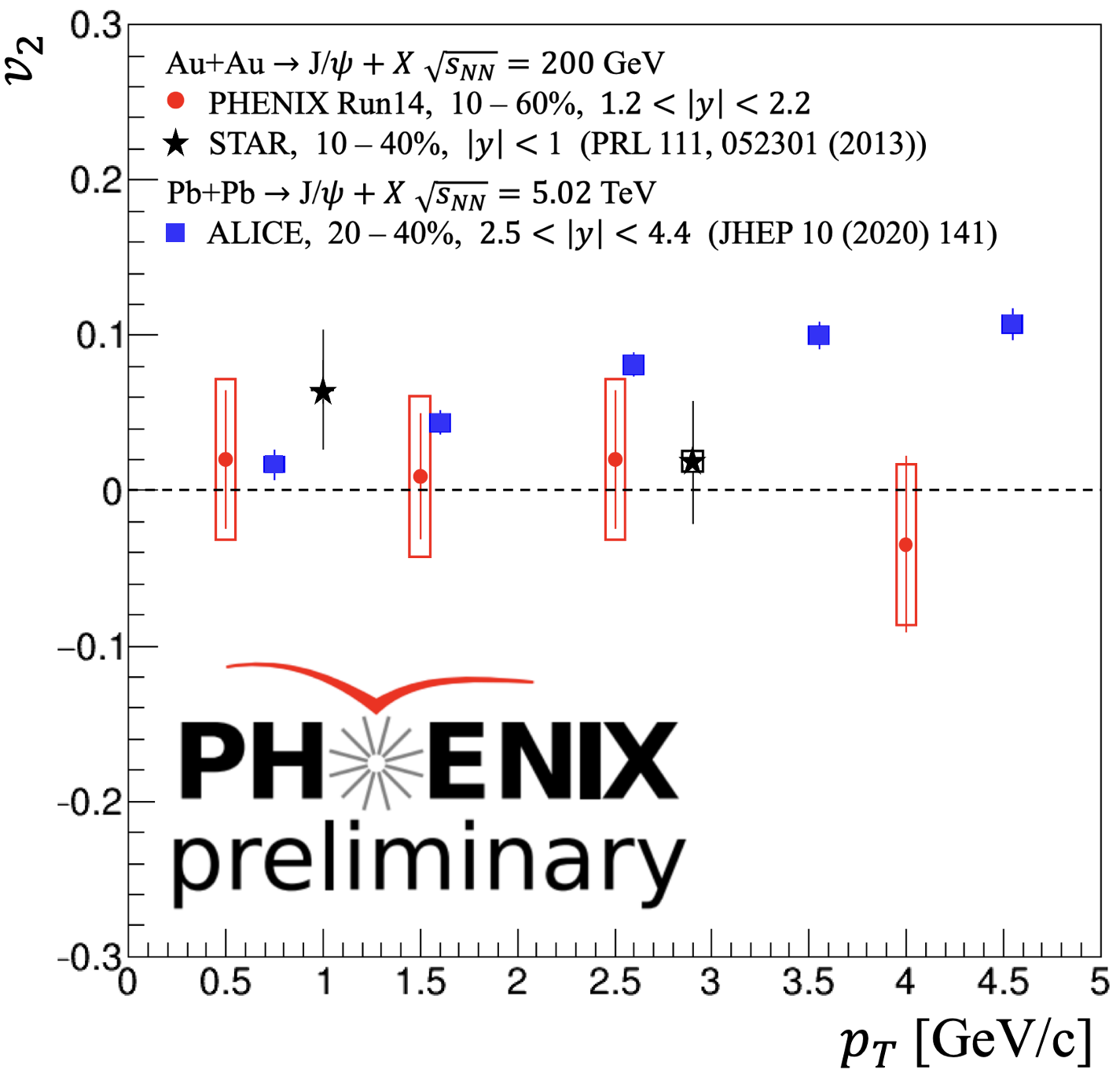}}
\caption{Plot of $p_T$ dependent $J/\psi$ $v_2$. The PHENIX result in light gray/red/circle is compared to STAR~\cite{Adamczyk_2013} in black/star and ALICE~\cite{2020} gray/blue/square.}
\label{Fig:F2H}
\end{figure}

From the figure we can see the clear nonzero $v_2$ measured by ALICE. Although the ALICE measurement is at a much higher energy, we know $v_2$ does not scale with energy for $J/\psi$, so it makes for a good comparison that the ALICE result which is clearly nonzero is different from our measurement. In our measurement, we see a $v_2$ that is clearly consistent with zero across all $p_T$ bins. The systematic uncertainties were conservatively estimated, not taking into account cancellations or correlations of uncertainties from different sources. Additional data from Run 16 of RHIC will be included in the final results, and we expect that both statistical and systematic uncertainties will be significantly reduced.

\section{Conclusion and Outlook}
We have presented PHENIX Run 14 $p_T$-dependent $J/\psi$ $v_2$ at forward rapidity at $\sqrt{s_{NN}}=200$ GeV. PHENIX has measured a $J/\psi$ $v_2$ that is consistent with zero. We have determined that the ALICE result, where there is clearly nonzero $v_2$, is distinctly different from our measurement, and that forward and mid-rapidity results at RHIC are consistent, but the uncertainties are still large. In the future, we will incorporate Run 16 data in our measurement, essentially doubling the current dataset and reducing statistical uncertainties accordingly. We also plan to study open heavy flavor $v_2$ to obtain a more complete understanding of the heavy flavor dynamics at RHIC.


\begin{thebibliography}{1}

\bibitem{Heinz_2009}
Ulrich Heinz.
\newblock The strongly coupled quark{\textendash}gluon plasma created at
  {RHIC}.
\newblock {\em Journal of Physics A: Mathematical and Theoretical},
  42(21):214003, May 2009.

\bibitem{Aidala_2014}
C.~Aidala et~al.
\newblock The {PHENIX} forward silicon vertex detector.
\newblock {\em Nuclear Instruments and Methods in Physics Research Section A:
  Accelerators, Spectrometers, Detectors and Associated Equipment}, 755:44--61,
  Aug 2014.

\bibitem{Poskanzer_1998}
A.~M. Poskanzer and S.~A. Voloshin.
\newblock Methods for analyzing anisotropic flow in relativistic nuclear
  collisions.
\newblock {\em Physical Review C}, 58(3):1671--1678, Sep 1998.

\bibitem{PhysRevC.84.054912}
A.~Adare et~al.
\newblock J/$\psi$ suppression at forward rapidity in {A}u+{A}u collisions at
  $\sqrt{{s}_{NN}}=200$ {GeV}.
\newblock {\em Physical Review C}, 84:054912, Nov 2011.

\bibitem{Andronic_2018}
Anton Andronic, Peter Braun-Munzinger, Krzysztof Redlich, and Johanna Stachel.
\newblock Decoding the phase structure of {QCD} via particle production at high
  energy.
\newblock {\em Nature}, 561(7723):321--330, Sep 2018.

\bibitem{Pereira_Da_Costa_2016}
H.~Pereira~Da Costa et~al.
\newblock Charmonium production in {P}b{\textendash}{P}b collisions with
  {ALICE} at the {LHC}.
\newblock {\em Nuclear Physics A}, 956:705--708, Dec 2016.

\bibitem{2020}
S.~Acharya et~al.
\newblock J/$\psi$ elliptic and triangular flow in {P}b-{P}b collisions at
  $\sqrt{s_{NN}}$ = 5.02 {TeV}.
\newblock {\em Journal of High Energy Physics}, 2020(10), Oct 2020.

\bibitem{Adamczyk_2013}
L.~Adamczyk et~al.
\newblock Measurement of {J}/$\psi$ {Azimuthal Anisotropy} in {A}u+{A}u
  {C}ollisions at $\sqrt{s_{NN}}=200$ {GeV}.
\newblock {\em Physical Review Letters}, 111(5), Aug 2013.

\end{thebibliography}

\end{document}